\documentclass[conference]{IEEEtran}
\IEEEoverridecommandlockouts

\usepackage{cite}
\usepackage{amsmath,amssymb,amsfonts}
\usepackage{algorithmic}
\usepackage{graphicx}
\usepackage{textcomp}
\usepackage{xcolor}

\usepackage{booktabs}%
\usepackage{multirow}

\def\BibTeX{{\rm B\kern-.05em{\sc i\kern-.025em b}\kern-.08em
    T\kern-.1667em\lower.7ex\hbox{E}\kern-.125emX}}
\begin{document}

\title{Multi-Modal Brain Tumor Segmentation via 3D Multi-Scale Self-attention and Cross-attention
}

\author{ \IEEEauthorblockN{Yonghao Huang\textsuperscript{1, 2}, Leiting Chen\textsuperscript{1,3}, Chuan Zhou\textsuperscript{1,3,}\footnotemark{*}}
\IEEEauthorblockA{\textsuperscript{1}Key Laboratory of Intelligent Digital Media Technology of Sichuan Province, \\University of Electronic Science and Technology of China\\
\textsuperscript{2}School of Computer Science and Engineering, University of Electronic Science and Technology of China\\
\textsuperscript{3}School of Information and Software Engineering, University of Electronic Science and Technology of China\\
huangyh@std.uestc.edu.cn, \{richardchen, zhouchuan\}@uestc.edu.cn}
}

\maketitle

\renewcommand{\thefootnote}{\fnsymbol{footnote}}
\footnotetext[1]{Chuan Zhou is the corresponding author.}

\begin{abstract}
  Due to the success of CNN-based and Transformer-based models in various computer vision tasks, recent works study the applicability of CNN-Transformer hybrid architecture models in 3D multi-modality medical segmentation tasks. Introducing Transformer brings long-range dependent information modeling ability in 3D medical images to hybrid models via the self-attention mechanism. However, these models usually employ fixed receptive fields of 3D volumetric features within each self-attention layer, ignoring the multi-scale volumetric lesion features. To address this issue, we propose a CNN-Transformer hybrid 3D medical image segmentation model, named TMA-TransBTS, based on an encoder-decoder structure. TMA-TransBTS realizes simultaneous extraction of multi-scale 3D features and modeling of long-distance dependencies by multi-scale division and aggregation of 3D tokens in a self-attention layer. Furthermore, TMA-TransBTS proposes a 3D multi-scale cross-attention module to establish a link between the encoder and the decoder for extracting rich volume representations by exploiting the mutual attention mechanism of cross-attention and multi-scale aggregation of 3D tokens. Extensive experimental results on three public 3D medical segmentation datasets show that TMA-TransBTS achieves higher averaged segmentation results than previous state-of-the-art CNN-based 3D methods and CNN-Transform hybrid 3D methods for the segmentation of 3D multi-modality brain tumors.
\end{abstract}


\section{Introduction}

\begin{figure}[htb]
    \begin{center}
        \includegraphics[width=0.49\textwidth]{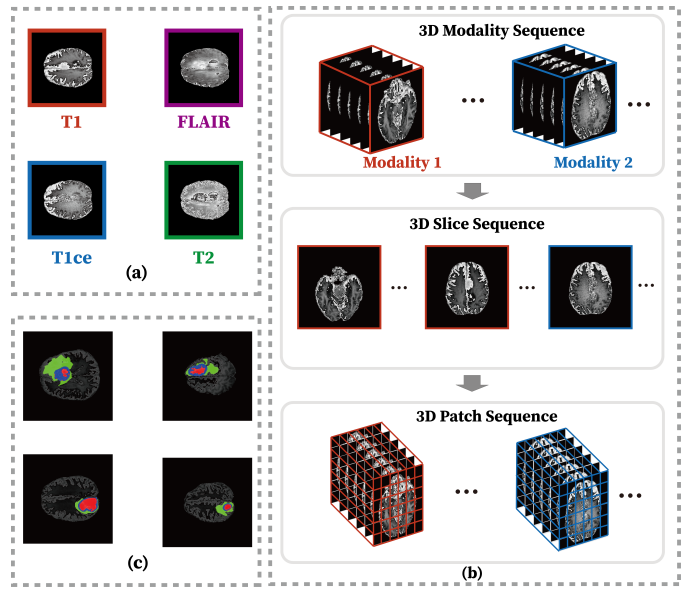}
        \caption{Schematic diagram of multi-modal brain MRI images. (a) different modalities of MRI images; (b) multi-modal MRI image data contain multiple sequence relationships; (c) brain tumors are characterized by high heterogeneity and high variability, and the red, green, and blue colors in the figure correspond to the segmentation labels of different categories of brain tumors, respectively}
        \label{Introduction}
    \end{center}
  \end{figure}

According to the reports, brain tumors, that is defined as an abnormal growth of tissues from any brain or skull component, constitute the 10th leading cause of death in developed countries and cause over 200,000 deaths every year \cite{2024A, Alessandra2013Living, The}. 
Magnetic Resonance Imaging (MRI) \cite{van1999basic} is the most common technique to diagnose brain cancers caused by brain tumors in clinical practice. 
As shown in Fig. \ref{Introduction} (a), multiple imaging modalities, including T1-weighted (T1), contrast-enhanced T1-weighted (T1ce), T2-weighted (T2), and Fluid Attenuated Inversion Recovery (FLAIR) images, are usually provided and used for detecting various tumor-induced tissues. 
The precise segmentation of tumor regions in multi-modal MRI scans can assist doctors in making accurate diagnosis conclusions and treatment plans. 
Thus, developing automatic and accurate multi-modal brain tumor segmentation techniques has received increased attention. 
Over the past decade, convolutional neural networks (CNNs) with an encoder-decoder architecture framework, such as UNet \cite{2015U}, have become dominant in the medical image segmentation field. 
Models like 3D UNet \cite{3DU-Net}, 3D Res-UNet \cite{3DU-Net}, V-Net \cite{milletari2016v}, ERV-Net \cite{zhou2021erv} and nnU-Net \cite{isensee2018nnu} 
have driven automated volumetric segmentation for various medical imaging modalities. 
However, it is difficult for 3D CNNs to model long-range dependencies and learn global semantic information due to the intrinsic locality of convolution operations. 

Medical imaging data, especially 3D multi-modal MRI imaging data, have many explicit sequence relationships (3D modal sequence, 3D slice sequence, and 3D patch sequence), which contain important multi-scale lesion information and long-distance dependencies, as shown in Fig. \ref{Introduction} (b). 
Specifically, multi-modal image sequences provided by different imaging techniques can provide modality-specific information, and this information can complement each other to provide a comprehensive basis for clinical diagnosis. 
Various imaging techniques differ in terms of color space, information density, etc. 
Tumors in MRI images from different modalities can present different appearances, so it is necessary to use multi-modal image data in combination to determine the location and size of the lesion area. 
For example, in brain tumor segmentation tasks, T1 MRI images and T1ce MRI images are suitable for detecting tumors without peritumoral edema, whereas T2 MRI images and FLAIR MRI images are sensitive to tumors with peritumoral edema \cite{zhou2019review}. 
Additionally, a single MRI image data is composed of multiple slices stacked on top of each other, and these sequential relationships provide information about the 3D structure of the detected site and the overall shape as well as the size of the lesion tissue. 
Additionally, as shown in Fig. \ref{Introduction} (c), since brain tumors are highly heterogeneous and highly variable, regions of the tumor may appear in various locations of the brain in different shapes and sizes and will exhibit a variety of biological characteristics. 
Ignoring these sequence relationships and the multi-scale lesion information will reduce the accuracy and reliability of brain tumor segmentation. 
Compared to CNNs, transformer-based methods \cite{ren2022shunted,dosovitskiy2020image,liu2021swin} have a global receptive field, which makes them able to learn the non-local interactions and long-range dependencies. 
Inspired by this, some works \cite{TransBTS,hatamizadeh2022unetr,shaker2024unetr++} have explored combining CNNs and Transformer and proposed CNN-Transformer hybrid models for various 3D medical image segmentation tasks. 
TransBTS \cite{TransBTS} for the first time introduces the Transformer into 3D CNNs for multi-modal brain tumor segmentation and achieves significant segmentation metrics improvement compared with previous 3D CNNs methods. 
However, the self-attention mechanism in the Transformer shows limitations for 3D multi-scale feature capture. 
More recently, nnFormer \cite{nnFormer} learns local and global feature representations in 3D volumes by jointly using volume-based local multi-head self-attention and volume-based global multi-head self-attention and provides rich receptive domain sizes. 
However, the receptive field size of nnFormer remains fixed in the same layer of self-attention, thereby leading to segmentation performance declining in handling multi-modal 3D MRI images with multiple brain tumors of different scales. 

To solve the above problems, we propose a brain tumor segmentation transformer based on 3D multi-scale attention, denoted TMA-TransBTS. 
TMA-TransBTS introduces a novel 3D multi-scale self-attention module (TMSM), which establishes different feature extraction branches in the self-attention mechanism through the shunted attention mechanism \cite{ren2022shunted}, divides and aggregates 3D markers with different granularities, and thus efficiently establishes long-distance dependencies in 3D multi-modal data and captures rich multi-scale features. 
Then, to provide richer contextual information for the decoding phase and help the network recover multi-scale spatial information in the decoding stage, a 3D Multi-scale Cross-attention Module (TMCM) is introduced to establish a link between the encoding and decoding phases. 
Finally, extensive experiments were conducted on three public 3D MRI brain tumor segmentation datasets, and the results showed that our model exhibited the best averaged segmentation results. 
By dividing the queried features in the cross-attention layer into different scales and calculating the corresponding cross-attention scores, this module is able to allow multi-scale information interaction and fusion between features from the encoder and those from the decoder in the same layer, further improving the segmentation performance of the model. 

In summary, our contributions are presented as follows: 
\begin{itemize} 
\item We propose a new brain tumor segmentation model based on 3D CNN-Transformer, which effectively eliminates the effects of lesion scale differences in multi-modal MRI images by establishing long-distance dependencies in 3D multi-scale features.
\item We propose a TMSM, which realizes simultaneous extraction of multi-scale 3D features and modeling of long-distance dependencies by multi-scale division and aggregation of 3D tokens in a self-attention layer. 
\item We propose a TMCM, which, by exploiting the mutual attention mechanism of cross-attention and multi-scale aggregation of 3D markers, can allow multi-scale information interaction and fusion of features from the encoder with those from the decoder in the same layer, which helps the network to recover multi-scale spatial information in the brain tumor and its surrounding tissues at the decoding stage and the preservation of detail information. 
\item We conducted extensive experiments on three publicly available brain tumor segmentation datasets. The experimental results show that compared to existing 3D medical segmentation models, TMA-TransBTS achieves better average segmentation results, which proves the effectiveness of the proposed method. 
\end{itemize}

\section{Related Work}
\subsection{CNN-based 3D Medical Image Segmentation Method}
Multi-modal medical images, such as MRI and CT, are stored in the form of 3D voxels. 
Applying convolution neural networks to process these 3D voxels for medical segmentation tasks, 3D U-Net \cite{3DU-Net} firstly extended the encoder-decoder architecture of 2D U-Net to 3D by employing 3D convolution kernels. 
Similar to 3D U-Net, V-Net \cite{milletari2016v} proposed an encoder-decoder architecture with 3D convolutions and to use the Dice Loss, to handle the imbalance between foreground and background voxels. 
nnU-Net \cite{isensee2018nnu} is proposed to focus on optimizing strategies of models, such as dataset processing, dynamic adaptation of network topologies, and inference strategies, instead of network architectures. 
Subsequently, nnU-Net was applied in diverse 3D medical image segmentation challenge competitions and achieved excellent results. 
ERV-Net \cite{zhou2021erv} introduced a post-processing method and a fusion loss function based Dice and Cross-entropy to improve the segmentation performance of 3D U-Net with residual blocks on the BraTS 2018 dataset. 
Recently, Liu et al. \cite{liu2023prior} designed a prior-based 3D U-Net with two-stage training strategies for knee-cartilage segmentation in MRI images. 
Different from these CNN-based 3D approaches, which find it difficult to build an explicit long-range dependence due to the inductive bias of convolution kernels, our Multi-scale UNETR inherits the superiority of transformers in global context modeling ability. 

\subsection{CNN-Transformer Hybrid 3D Medical Image Segmentation Method}
Thanks to the long-range dependency modeling ability of transformers, transformer-based frameworks have gained popularity and achieved performance improvement on various computer vision tasks, including classification \cite{dosovitskiy2020image, liu2021swin, han2021transformer}, segmentation \cite{cao2022swin, karimi2021convolution}, and detection \cite{hassani2023neighborhood, carion2020end}. 
Inspired by these works, the recent works \cite{TransBTS, li2022transbtsv2, li2023attention, hatamizadeh2022unetr, hatamizadeh2021swin, shaker2024unetr++, nnFormer, chen2021transunet, lin2022ds, valanarasu2021medical, zhang2021transfuse} have explored 3D CNN-Transformer hybrid architectures to combine 3D convolutions and self-attention operations for better 3D medical image segmentation. 
TransBTS \cite{TransBTS} presented the first attempt to merge 3D CNNs and transformers for 3D multi-modal brain tumor segmentation tasks. 
TransBTSV2 \cite{li2022transbtsv2} introduced deformable convolution in the skip-connection part of TransBTS and extended to general 3D medical segmentation tasks, including the brain tumor segmentation task, the liver tumor segmentation task and the kidney tumor segmentation task. 
Li et al. \cite{li2023attention} proposed an 3D CBAM to enhance the irrgular lesion features extracting ability of CNN-Transformer hybrid networks. 
UNETR \cite{hatamizadeh2022unetr} built a U-shape CNN-Transformer 3D segmentation network by utilizing transformers as the encoder to extract global context features and CNNs as the decoder for feature decoding. 
Swin UNETR \cite{hatamizadeh2021swin} employed transformers based on shifted windows to extract features of different resolutions. 
UNETR++ \cite{shaker2024unetr++} proposed an efficient paired attention block to construct a lightweight CNN-Transformer hybrid model for various 3D medical image segmentation tasks. 
To more efficiently combine convolution and self-attention operations for improving 3D medical image segmentation performance, nnFormer \cite{nnFormer} proposed combining interleaved 3D convolution and self-attention operations. 
Additionally, nnFormer jointly used local and global volume-based self-attention to extract local and global 3D features and model inter-slice long-range dependencies. 

\section{Methods}
\begin{figure*}[htb]
    \begin{center}
        \includegraphics[width=1.00\textwidth]{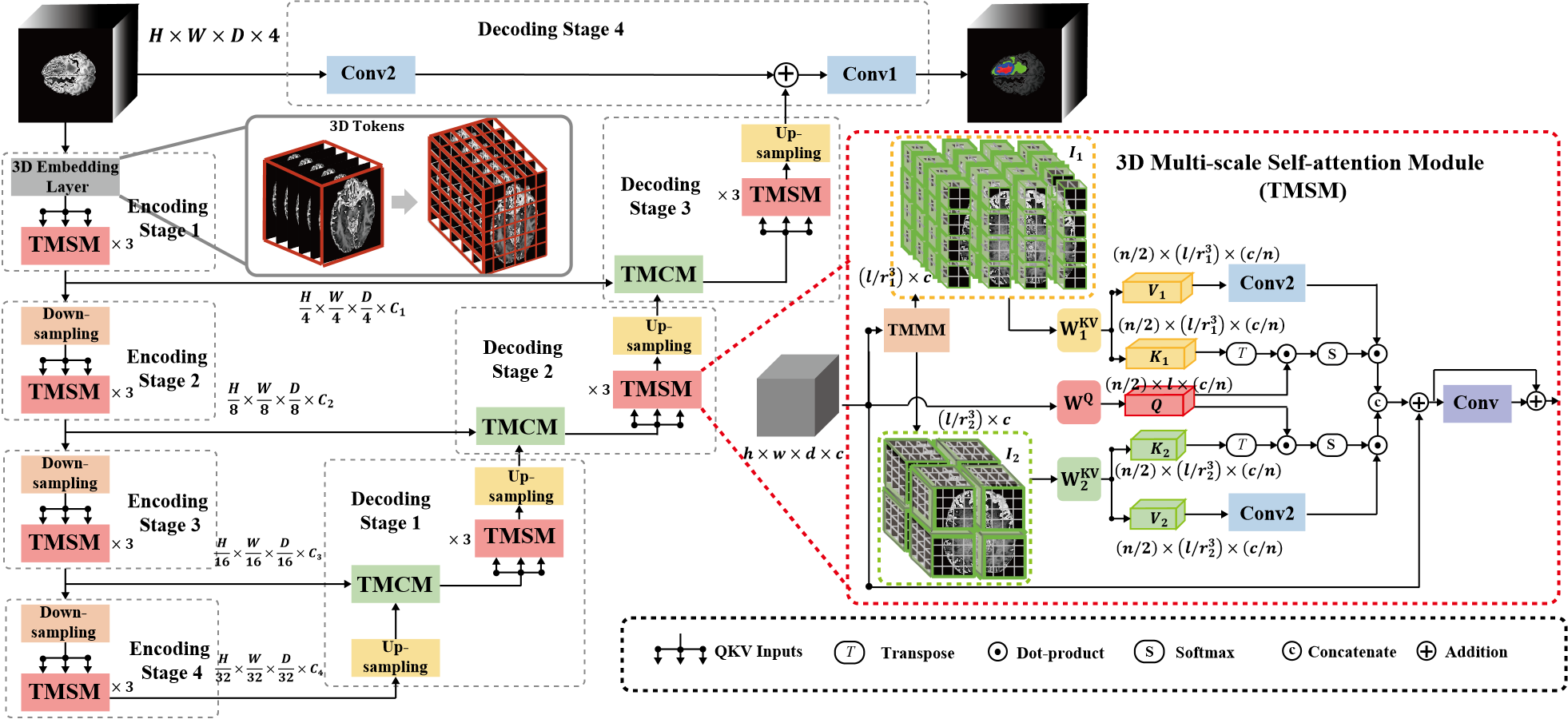}
        \caption{Architecture of TMA-TransBTS. TMA-TransBTS contains four encoding stages and four decoding stages. }
        \label{Methods1}
    \end{center}
  \end{figure*}
\begin{figure*}[htb]
    \begin{center}
        \includegraphics[width=0.8\textwidth]{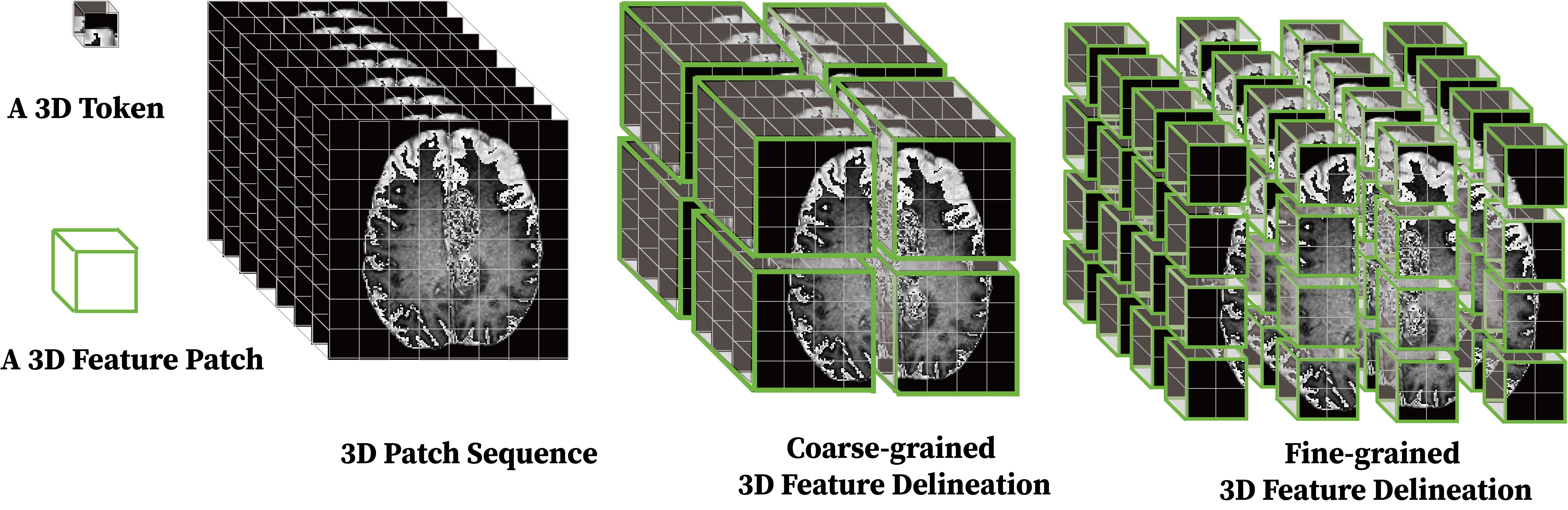}
        \caption{Multi-scale feature delineation by varying degrees of aggregation of 3D tokens. }
        \label{multiscale}
    \end{center}
\end{figure*}

\subsection{Overall Architecture}
An overview of the proposed TMA-TransBTS is presented in Fig. \ref{Methods1}. 
TMA-TransBTS is a network of encoder-decoder structures. 
The encoder involves four encoding stages for modeling long-range dependencies and extracting multi-scale features in multi-modality brain MRI images. 
The input of TMA-TransBTS is a 3D patch $\mathbf{X} \in \mathbb{R}^{H \times W \times D \times C}$, where $H$, $W$, $D$, and $C$ denote the height, width, depth of each input scan and the number of modalities, respectively. 
In this work, $C = 4$. 
The encoding stage 1 contains a 3D patch embedding layer and three TMSM layers. 
The output feature $\mathbf{F}_{en1} \in \mathbb{R}^{\frac{H}{4} \times \frac{W}{4} \times \frac{D}{4} \times C_{1}}$ of the encoding stage 1 can be formulated as:
\begin{flalign}\label{function1}
  \mathbf{F}_{en1}=\operatorname{TMSM}\left(\operatorname{conv}_{4 \times 4 \times 4}(\mathrm{\mathbf{X}})\right)
\end{flalign}
where $\operatorname{conv}_{4 \times 4 \times 4}(\cdot)$ is a $4 \times 4 \times 4$ 3D convolution with stride 4, and $\operatorname{TMSM}(\cdot)$ denotes three TMSM layers. 
Each convolution operation contains a Parametric Rectified Linear Unit \cite{PReLU} function and a Batch Normalization. 
The remaining three encoding stages all contain a downsampling layer and three TMSM layers. 
We utilize the downsampling layer to decrease the resolution by using non-overlapping convolution with a factor two. 
The output features $\mathbf{F}_{en2} \in \mathbb{R}^{\frac{H}{4} \times \frac{W}{4} \times \frac{D}{4} \times C_{1}}$, $\mathbf{F}_{en3} \in \mathbb{R}^{\frac{H}{4} \times \frac{W}{4} \times \frac{D}{4} \times C_{1}}$, and $\mathbf{F}_{en4} \in \mathbb{R}^{\frac{H}{4} \times \frac{W}{4} \times \frac{D}{4} \times C_{1}}$ of the encoding stage 2, the encoding stage 3, and the encoding stage 4 can be formulated as:
\begin{flalign}
\begin{split}
    & \mathbf{F}_{e n 2}=\operatorname{TMSM}\left(\operatorname{conv}_{2 \times 2 \times 2}\left(\mathbf{F}_{e n 1}\right)\right) \\
    & \mathbf{F}_{e n 3}=\operatorname{TMSM}\left(\operatorname{conv}_{2 \times 2 \times 2}\left(\mathbf{F}_{e n 2}\right)\right) \\
    & \mathbf{F}_{e n 4}=\operatorname{TMSM}\left(\operatorname{conv}_{2 \times 2 \times 2}\left(\mathbf{F}_{e n 3}\right)\right)
\end{split}
\end{flalign}
where $\operatorname{conv}_{2 \times 2 \times 2}(\cdot)$ is a $2 \times 2 \times 2$ 3D convolution with stride 2. In this work, $c_{1}$, $c_{2}$, $c_{3}$ and $c_{4}$ are set to 32, 64, 128, 256, respectively. 

The decoder comprises four stages and accepts the multi-scale features from the encoder. 
Skip connections in conventional segmentation models \cite{3DU-Net} are often pixel-by-pixel summing operations or concatenation operations. 
Inspired by the skip attention in \cite{nnFormer}, we propose a TMCM to replace the conventional skip connections. 
TMCM not only effectively preserves the information in the encoder, but also allows the decoder to focus on features at different scales. 
In the decoder, the decoding stage 1 contains two up-sampling layers, a TMCM, and three TMSM layers. 
The output of decoding stage 1 denotes $\mathbf{F}_{de1} \in \mathbb{R}^{\frac{H}{8} \times \frac{W}{8} \times \frac{D}{8} \times C_{2}}$. 
The outputs of decoding stage 2, decoding stage 3, and decoding stage 4 can be formulated as:

\begin{flalign}
    \begin{split}
    & \mathbf{F}_{de2}=\operatorname{tconv_{2}}\left(\operatorname{TMSM}\left(\operatorname{TMCM}\left(\mathbf{F}_{en2}, \mathbf{F}_{de1}\right)\right)\right) \\
    & \mathbf{F}_{de3}=\operatorname{tconv_{2}}\left(\operatorname{TMSM}\left(\operatorname{TMCM}\left(\mathbf{F}_{en1}, \mathbf{F}_{de2}\right)\right)\right) \\
    & \mathbf{F}_{de4}=\operatorname{Conv1}\left(\operatorname{Conv2}\left(\mathbf{X}\right) + \mathbf{F}_{de4}\right)
    \end{split}
    \end{flalign}
  
where $\operatorname{tconv_{2}}$ is a transposed convolution with a upsampling factor 2. 
$\operatorname{Conv1}$ contains a $3 \times 3 \times 3$ residual convolution block and a $1 \times 1 \times 1$  convolution block. 
$\operatorname{Conv2}$ is a $3 \times 3 \times 3$ residual convolution block. 
The deep supervised stage accepts the feature vectors from the four decoding stages and computes the loss function between each of these four feature vectors and the true label of the corresponding resolution. 

\subsection{3D Multi-Scale Self-Attention Module}

The proposed TMSM maps 3D features to different scale spaces by an efficient 3D multi-scale map module (TMMM) and effectively captures enriched multi-scale feature representations. 
TMSM uses shared query vectors to calculate self-attention weights with key and value vectors of different scales. 
The architecture of TMSM is shown in Fig. \ref{Methods1}. 
As shown in Fig. \label{multiscale}, in the coarse-grained feature extraction branch, the TMSM aggregates a large number of 3D tokens to achieve the learning of large-scale lesion features and to reduce the computational effort; in the fine-grained feature extraction branch, the TMSM aggregates a small number of 3D tokens to retain the model's ability to focus on small-scale lesion features. 
Given an input tensor $\mathbf{I} \in \mathbb{R}^{hwd \times c}$, we compute the shared query tensor $\mathbf{Q}_{shared} \in \mathbb{R}^{hwd \times c}$ with the projection weight $\mathbf{W}^{Q}$, where $\mathbf{Q}_{shared} = \mathbf{W}^{Q} \mathbf{I}$. 
TMMM, which contains two parallel depthwise convolution layers \cite{depthwise}, two reshape operations, two layer normalization (LN) \cite{LN} and two GELU \cite{GELU} functions, maps $\mathbf{I}$ to two different scale spaces. 
The kernel sizes of the two parallel depthwise convolution layers denote $r_{1}$ and $r_{2}$, respectively. 
We set four TMMM variants with different parameters $r_{1}$ and $r_{2}$ to model long-range dependencies in lesion features of different scales. 
In encoding stage 1 and decoding stage 3, we set $r_{1} = 8$ and $r_{2} = 4$. 
In encoding stage 2 and decoding stage 2, we set $r_{1} = 4$ and $r_{2} = 2$. 
In encoding stage 3 and decoding stage 1, we set $r_{1} = 2$ and $r_{2} = 1$. 
Finally, we set $r_{1} = r_{2} = 1$ in encoding stage 4. 
We denote the outputs of TMMM as $\mathbf{I}_{1} \in \mathbb{R}^{(l / r_{1}^{3}) \times c}$ and $\mathbf{I}_{2} \in \mathbb{R}^{(l / r_{1}^{3}) \times c}$, respectively, where $l = hwd$. 
Then, we map $\mathbf{I}_{1}$ into the key and value space with a linear tensor $\mathbf{W}_{1}^{KV}$ and divide the result into key tensor $\mathbf{K}_{1}$ of shape $\frac{n}{2} \times \frac{l}{r_{1}^{3}} \times \frac{c}{n}$ and value tensor $\mathbf{V}_{1}$ of shape $\frac{n}{2} \times \frac{l}{r_{1}^{3}} \times \frac{c}{n}$. 
Similarly, we map $\mathbf{I}_{2}$ into the key and value space with a linear tensor $\mathbf{W}_{1}^{KV}$ and divide the result into key tensor $\mathbf{K}_{1}$ of shape $\frac{n}{2} \times \frac{l}{r_{1}^{3}} \times \frac{c}{n}$ and value tensor $\mathbf{V}_{1}$ of shape $\frac{n}{2} \times \frac{l}{r_{1}^{3}} \times \frac{c}{n}$. 
It's worth noting that $n$ is the total multi-head number in the multi-head self-attention computing process. 
Finally, the shared query tensor $\mathbf{Q}_{shared}$ is divided into $\mathbf{Q}_{1}$ of shape $\frac{n}{2} \times l \times \frac{c}{n}$ and $\mathbf{Q}_{2}$ of shape $\frac{n}{2} \times l \times \frac{c}{n}$. 
The attention results of different scale spaces are defined as follows: 
  
\begin{flalign}
\begin{split}
      & \mathbf{X}_1=\operatorname{Softmax}\left(\frac{\mathbf{Q}_1 \mathbf{K}_1^T}{\sqrt{d_1}}\right) \cdot \operatorname{Conv2}\left(\mathbf{V}_1\right) \\
      & \mathbf{X}_2=\operatorname{Softmax}\left(\frac{\mathbf{Q}_2 \mathbf{K}_2^T}{\sqrt{d_2}}\right) \cdot \operatorname{Conv2}\left(\mathbf{V}_2\right)
\end{split}
\end{flalign}
  
where $d_{1}$ and $d_{2}$ are the size of corresponding tensor, and $\operatorname{Softmax}(\cdot)$ is the softmax function. 
Subsequently, $\mathbf{X}_{1}$ and $\mathbf{X}_{2}$ are concatenated to obtain the multi-scale attention tensor. 
Finally, a residual connection and a convolution block are employed to further extract rich feature information. 
The convolution block contains a $3 \times 3 \times$ convolution block and a $1 \times 1 \times 1$. 

\subsection{3D Multi-Scale Cross-Attention Module}
The skip connection is an important part of segmentation networks based on encoder-decoder architecture. 
Conventional skip connections in previous segmentation networks usually adapt either concatenation or summation. 
Inspired by the skip attention in nnFormer \cite{nnFormer}, we proposed to replace the conventional skip connections with a cross-attention mechanism, which is named TMCM in this paper. 
On the one hand, TMCM maps the input tensor into key tensor and value tensor, respectively. 
On the other hand, TMCM in decoding stages maps the output of the symmetrical encoding stage into query tensor. 
For instance, TMCM in the decoding stage 3 accepts the output of the encoding stage 1, $F_{en1}$, and the output of the decoding stage 2, $F_{de2}$, as the input. 
Similar to TMSM, TMCM maps $\mathbf{F}_{en1}$ into a shared query tensor and divides the query tensor into $\hat{\mathbf{Q}}_{1}$ and $\hat{\mathbf{Q}}_{2}$ according to the multi-head number $n$. 
Then, TMSM employ a TMMM to map $\mathbf{F}_{de2}$ into two feature maps of different scales, i.e., $\hat{\mathbf{I}}_{1}$ and $\hat{\mathbf{I}}_{2}$. 
In the decoding stage 1, we set $r_{1} = 2$ and $r_{2} = 1$. 
In the decoding stage 2, we set $r_{1} = 4$ and $r_{2} = 2$. 
In the decoding stage 3, we set $r_{1} = 8$ and $r_{2} = 4$. 
Subsequently, two map matrices are employed to map $\hat{\mathbf{I}}_{1}$ and $\hat{\mathbf{I}}_{2}$ into two key and value spaces, respectively. 
We denote the key tensor and value tensor of $\hat{\mathbf{I}}_{1}$ as $\hat{\mathbf{K}}_{1}$ and $\hat{\mathbf{V}}_{1}$, respectively. 
Similarly, we denote the key tensor and value tensor of $\hat{\mathbf{I}}_{2}$ as $\hat{\mathbf{K}}_{2}$ and $\hat{\mathbf{V}}_{2}$, respectively. 
The cross-attention results of different scale spaces are defined as follows: 
\begin{flalign}
\begin{split}
        & \hat{\mathbf{X}}_1=\operatorname{Softmax}\left(\frac{\hat{\mathbf{Q}}_1 \hat{\mathbf{K}}_1^T}{\sqrt{d_1}}\right) \cdot \operatorname{Conv2}\left(\hat{\mathbf{V}}_1\right) \\
        & \hat{\mathbf{X}}_2=\operatorname{Softmax}\left(\frac{\hat{\mathbf{Q}}_2 \hat{\mathbf{K}}_2^T}{\sqrt{d_2}}\right) \cdot \operatorname{Conv2}\left(\hat{\mathbf{V}}_2\right)
\end{split}
\end{flalign}
Finally, TMCM uses the same residual connection and convolution block as TMSM to further extract the commanded lesion features. 
  
\subsection{Deep Supervision Stage and Loss Function}
We add a deep supervision stage during the training stage to help the model learning more information following in \cite{nnFormer}. 
Specifically, the feature of each decoding stage is used for computing dice loss. 
We denote the segmentation results from the decoding stage 1, the decoding stage 2, the decoding stage 3 and the decoding stage 4 as $\mathbf{X} \in \mathbb{R}^{\frac{H}{16} \times \frac{W}{16} \times \frac{D}{16} \times 3}$, $\mathbf{Y} \in \mathbb{R}^{\frac{H}{8} \times \frac{W}{8} \times \frac{D}{8} \times 3}$, $\mathbf{Z} \in \mathbb{R}^{\frac{H}{4} \times \frac{W}{4} \times \frac{D}{4} \times 3}$ and $\mathbf{W} \in \mathbb{R}^{H \times W \times D \times 3}$, respectively. 
We down-sample the ground truth segmentation mask to match these segmentation results' resolution. 
These downsampling ground truth segmentation masks are denoted as $\hat{\mathbf{X}} \in \mathbb{R}^{\frac{H}{16} \times \frac{W}{16} \times \frac{D}{16} \times 3}$, $\hat{\mathbf{Y}} \in \mathbb{R}^{\frac{H}{8} \times \frac{W}{8} \times \frac{D}{8} \times 3}$, $\hat{\mathbf{Z}} \in \mathbb{R}^{\frac{H}{4} \times \frac{W}{4} \times \frac{D}{4} \times 3}$ and $\hat{\mathbf{W}} \in \mathbb{R}^{H \times W \times D \times 3}$. 
For the segmentation result in each stage, we set a dice loss function: 
  
\begin{flalign}
\begin{split}
    \begin{aligned}
        & \mathcal L_1=1-\sum_{i=1}^I\left(\frac{2 * \sum_{n=1}^{N_1} X_{n, i} \cdot \hat{X}_{n, i}}{\sum_{n=1}^{N_1} X_{n, i}^2+\sum_{n=1}^{N_1} \hat{X}_{n, i}^2}\right) \\
        & \mathcal L_2=1-\sum_{i=1}^I\left(\frac{2 * \sum_{n=1}^{N_2} Y_{n, i} \cdot \hat{Y}_{n, i}}{\sum_{n=1}^{N_2} Y_{n, i}^2+\sum_{n=1}^{N_2} \hat{Y}_{n, i}^2}\right) \\
        & \mathcal L_3=1-\sum_{i=1}^I\left(\frac{2 * \sum_{n=1}^{N_3} Z_{n, i} \cdot \hat{Z}_{n, i}}{\sum_{n=1}^{N_3} Z_{n, i}^2+\sum_{n=1}^{N_3} \hat{Z}_{n, i}^2}\right) \\
        & \mathcal L_4=1-\sum_{i=1}^I\left(\frac{2 * \sum_{n=1}^{N_4} W_{n, i} \cdot \hat{W}_{n, i}}{\sum_{n=1}^{N_4} W_{n, i}^2+\sum_{n=1}^{N_4} \hat{W}_{n, i}^2}\right)
    \end{aligned}
\end{split}
\end{flalign}
  
where $I$ denotes the number of targets; $N_{1}$, $N_{2}$, $N_{3}$ and $N_{4}$ denote the number of voxels; $X_{n, i}$, $Y_{n, i}$, $Z_{n, i}$ and $W_{n, i}$ are output results at voxel $n$ for class $i$; $\hat{X}_{n, i}$, $\hat{Y}_{n, i}$, $\hat{Z}_{n, i}$ and $\hat{W}_{n, i}$ are the ground truths at voxel $n$ for class $i$. 
  
The final loss can be formulated as follows: 
  
\begin{flalign}
  \mathcal L_{a l l}=\alpha_1 \mathcal L_1+\alpha_2 \mathcal L_2+\alpha_3 \mathcal L_3+\alpha_4 \mathcal L_4
\end{flalign}
  
where $\alpha_{1}$, $\alpha_{2}$, $\alpha_{3}$, and $\alpha_{4}$ are set to 0.125, 0.25, 0.5, and 1.0 according to different resolutions, respectively. 

\section{Experiments}

\begin{table*}[htbp]
  \centering
  \caption{Comparison on BraTS 2018 dataset. The evaluation metrics are HD (mm) and DSC scores (\%). Best results are bolded while second best are underline. }
    \begin{tabular}{l|c|c|c|c|c|c|c|c|c|c}
    \toprule
    \multicolumn{1}{c|}{\multirow{2}[4]{*}{Methods}} & \multicolumn{4}{c}{Dice (\%)}  & \multicolumn{4}{c|}{HD (mm)}   & \multirow{2}[4]{*}{Params (M)} & \multirow{2}[4]{*}{FLOPs (G)} \\
\cmidrule{2-9}          & ET    & TC    & WT    & Average & ET    & TC    & WT    & Average &       &  \\
    \midrule
    3D U-Net \cite{3DU-Net}& \underline{73.88}  & \underline{81.32}  & 89.71  & \underline{81.64}  & 4.765  & \underline{6.645}  & 7.568  & \underline{6.326}  & \textbf{16.32} & 1899.47 \\
    Residual 3D U-Net \cite{yu2019liver}& 72.80  & 79.87  & \underline{89.81}  & 80.83  & 4.705  & 9.342  & \textbf{6.890}  & 6.979  & 113.74 & 2601.05 \\
    TransBTS w/o TTA \cite{TransBTS}& 69.36  & 77.47  & 87.09  & 77.97  & 10.240  & 17.004  & 16.394  & 14.546  & 32.99 & 263.73 \\
    TransBTS w TTA \cite{TransBTS}& 71.05  & 81.20  & 88.07  & 80.11  & 7.328  & 10.111  & 13.429  & 10.289  & 32.99 & 263.73 \\
    UNETR \cite{hatamizadeh2022unetr}& 71.07  & 78.09  & 87.97  & 79.04  & 6.211  & 11.096  & 10.720  & 9.342  & 102.12 & 179.78 \\
    Swin UNETR \cite{hatamizadeh2021swin}& 70.71  & 74.24  & 88.21  & 77.72  & \textbf{4.349}  & 14.993  & 9.060  & 9.467  & 62.19 & 794.02 \\
    UNETR++ \cite{shaker2024unetr++}& 71.05  & 81.20  & 88.07  & 80.11  & 7.328  & 10.111  & 13.429  & 10.289  & 42.65 & \textbf{139.41} \\
    nnFormer \cite{nnFormer}& 70.17  & 79.76  & 89.26  & 79.73  & 4.983  & 9.572  & 7.231  & 7.262  & 149.68 & 372.06 \\
    \midrule
    TMA-TransBTS & \textbf{74.30}  & \textbf{82.40}  & \textbf{90.38}  & \textbf{82.36}  & \underline{4.592}  & \textbf{6.285}  & \underline{7.134}  & \textbf{6.004}  & \underline{30.85} & \underline{141.79} \\
    \bottomrule
    \end{tabular}%
  \label{BraTS2018}%
\end{table*}%

\begin{table*}[htbp]
  \centering
  \caption{Comparison on BraTS 2019 dataset. The evaluation metrics are HD (mm) and DSC scores (\%). Best results are bolded while second best are underline. }
    \begin{tabular}{l|c|c|c|c|c|c|c|c|c|c}
    \toprule
    \multicolumn{1}{c|}{\multirow{2}[4]{*}{Methods}} & \multicolumn{4}{c}{Dice (\%)}  & \multicolumn{4}{c|}{HD (mm)}   & \multirow{2}[4]{*}{Params (M)} & \multirow{2}[4]{*}{FLOPs (G)} \\
\cmidrule{2-9}          & ET    & TC    & WT    & Average & ET    & TC    & WT    & Average &       &  \\
    \midrule
    3D U-Net \cite{3DU-Net}& 69.33  & 77.36  & 88.61  & 78.43  & 14.319  & 21.189  & 20.504  & 18.671  & \textbf{16.32} & 1899.47 \\
    Residual 3D U-Net \cite{yu2019liver}& 69.41  & 76.55  & 89.46  & 78.47  & 10.951  & 14.018  & 14.455  & 13.141  & 113.74 & 2601.05 \\
    TransBTS w/o TTA \cite{TransBTS}& 69.84  & 76.13  & 88.43  & 78.13  & 12.343  & 14.534  & 17.392  & 14.756  & 32.99 & 263.73 \\
    TransBTS w TTA \cite{TransBTS}& 70.73  & 76.99  & 90.10  & 79.27  & 11.725  & 13.165  & 10.818  & 11.903  & 32.99 & 263.73 \\
    UNETR \cite{hatamizadeh2022unetr}& 70.57  & 74.22  & 89.53  & 78.11  & \underline{10.136}  & 12.985  & 11.066  & \underline{11.396}  & 102.12 & 179.78 \\
    Swin UNETR \cite{hatamizadeh2021swin}& \underline{71.60}  & \textbf{79.61} & 88.82 & \underline{80.01}  & 12.129 & \underline{12.541} & 20.428 & 15.033  & 62.19 & 794.02 \\
    UNETR++ \cite{shaker2024unetr++}& 68.57  & 76.08  & \underline{90.81}  & 78.49  & 13.822  & 13.679  & \underline{8.346}  & 11.949  & 42.65 & \textbf{139.41} \\
    nnFormer \cite{nnFormer}& 67.59  & 75.04  & 87.86  & 76.83  & 11.993  & 14.663  & 14.304  & 13.653  & 149.68 & 372.06 \\
    \midrule
    TMA-TransBTS & \textbf{72.52}  & \underline{78.73}  & \textbf{91.42}  & \textbf{80.89}  & \textbf{9.138}  & \textbf{9.401}  & \textbf{5.697}  & \textbf{8.079}  & \underline{30.85} & \underline{141.79} \\
    \bottomrule
    \end{tabular}%
  \label{BraTS2019}%
\end{table*}%

\begin{table*}[htbp]
  \centering
  \caption{Comparison on BraTS 2020 dataset. The evaluation metrics are HD (mm) and DSC scores (\%). Best results are bolded while second best are underline. }
    \begin{tabular}{l|c|c|c|c|c|c|c|c|c|c}
    \toprule
    \multicolumn{1}{c|}{\multirow{2}[4]{*}{Methods}} & \multicolumn{4}{c}{Dice (\%)}  & \multicolumn{4}{c|}{HD (mm)}   & \multirow{2}[4]{*}{Params (M)} & \multirow{2}[4]{*}{FLOPs (G)} \\
\cmidrule{2-9}          & ET    & TC    & WT    & Average & ET    & TC    & WT    & Average &       &  \\
    \midrule
    3D U-Net \cite{3DU-Net}& 72.49  & 78.47  & 90.74  & 80.57  & 6.439  & 8.936  & \underline{5.574}  & 6.983  & \textbf{16.32} & 1899.47 \\
    Residual 3D U-Net \cite{yu2019liver}& 73.72  & 78.43  & 90.10  & 80.75  & 6.499  & 9.935  & 8.043  & 8.159  & 113.74 & 2601.05 \\
    TransBTS w/o TTA \cite{TransBTS}& 71.07  & 77.02  & 89.59  & 79.23    & 8.960    & 11.150    & 11.370    & 8.485  & 32.99 & 263.73 \\
    TransBTS w TTA \cite{TransBTS}& 73.09  & 79.20    & 91.12    & 81.14    & 7.400    & 8.410    & 7.950    & 7.230  & 32.99 & 263.73 \\
    UNETR \cite{hatamizadeh2022unetr}& 69.97  & 76.60  & 88.50  & 78.36  & 8.475  & 14.937  & 18.771  & 14.061  & 102.12 & 179.78 \\
    SwinUNETR \cite{hatamizadeh2021swin}& 72.18  & \underline{79.59}  & \underline{91.87}  & \underline{81.21}  & \textbf{5.221}  & 9.738  & 5.706  & 6.888  & 62.19 & 794.02 \\
    UNETR++ \cite{shaker2024unetr++}& \underline{74.30}  & 77.49  & 91.03  & 80.94  & 6.019  & \underline{8.019}  & 6.209  & \underline{6.749}  & 42.65 & \textbf{139.41} \\
    nnFormer \cite{nnFormer}& 71.45  & 79.57  & 91.80  & 80.94  & 6.800  & 8.823  & 6.989  & 7.537  & 149.68 & 372.06 \\
    \midrule
    TMA-TransBTS & \textbf{74.86}  & \textbf{79.64}  & \textbf{92.31}  & \textbf{82.27}  & \underline{5.904}  & \textbf{6.603}  & \textbf{4.537}  & \textbf{5.681}  & \underline{30.85} & \underline{141.79} \\
    \bottomrule
    \end{tabular}%
  \label{BraTS2020}%
\end{table*}%

\subsection{Dataset and Evaluation Metrics}
For thoroughly comparing TMA-TransBTS to previous CNN-based 3D medical segmentation methods and CNN-Transformer hybrid 3D medical image segmentation methods, we conduct experiments on three datasets: BraTS 2018 dataset, BraTS 2019 dataset and BraTS 2020 dataset. Specifically, the three datasets are provided by the brain tumor segmentation challenge \cite{BraTs1, BraTs2, BraTs3}. BraTS 2018 dataset contains training set and validation set. The training set contains 285 3D MRI images and the validation set contains 66 3D MRI images. 
Each sample contains four modalities of brain MRI scans, namely T1, T1ce, T2 and FLAIR. 
The size of all MRI images is 240 $\times$ 240 $\times$ 155.
The target categories contain Whole Tumor (WT), Enhancing Tumor (ET) and Tumor Core (TC).  
These target categories are composed of some tumor sub-regions, namely necrotic, non-enhancing tumor, peritumoral edema and enhancing tumor. 
The BraTS 2019 dataset consists of 335 cases for training and 125 cases for validation. 
The BraTS 2020 dataset consists of 369 cases for training and 125 cases for validation. 
Except for the number of samples, the other information about these two datasets are the same. 
The validation sets of the three datasets do not provide labels, so we split the training sets into 7 : 1 : 2 ratio for training, validation and testing and report on the test set. 
We use Dice Score (DSC) and 95\% Hausdorff Distance (HD) on three brain tumor categories, which are WT, ET and TC. 
Besides, we also report the average DSC and HD on the three brain tumor categories. 

\subsection{Experiment details}
We run all experiments based on Python 3.9.19, PyTorch 2.4.1 and Ubuntu 20.04. 
All models are trained with a NVIDIA RTX 4090 for 500 epochs from scratch using a batch size of 2. 
The default optimizer is Adam. 
With warm-up strategy for 30 epochs during training,the initial learning rate is set to 0.0001 with a cosine learning rate decay schedule. 
We set $n = 16$ in our model. 
We use conventional strategies for image augmentation including random mirror flip, random cropping the data from 240 $\times$ 240 $\times$ 155 to 128 $\times$ 128 $\times$ 128 voxels and random intensity shift. 
We conducted three repeated experiments, and the final result is the average of the three experiments. 

\subsection{Main Results}
We make experiments on BraTS 2018, BraTS 2019, and BraTS 2020 datasets to compare our TMA-TransBTS against CNN-based 3D methods, including 3D U-Net \cite{3DU-Net}, Residual 3D U-Net \cite{yu2019liver} and the state of the art CNN-Transformer Hybrid 3D methods, including TransBTS \cite{TransBTS}, UNETR \cite{hatamizadeh2022unetr}, SwinUNETR \cite{hatamizadeh2021swin}, UNETR++ \cite{shaker2024unetr++}, and nnFormer \cite{nnFormer}. 

\textbf{BraTS 2018.} Our TMA-TransBTS achieve the highest DSC and the lowest HD in all brain tumour classes. Moreover, TMA-TransBTS is able to surpass the second best approach in DSC and HD, i.e., 3D U-Net. For instance, TMA-TransBTS outperforms 3D U-Net by 0.72\% in average DSC and 0.322 mm average HD. In comparison to previous CNN-based 3D approaches and CNN-Transformer Hybrid 3D approaches, TMA-TransBTS shows significant improvements in most classes. 

\textbf{BraTS 2019.} We also evaluate TMA-TransBTS on the BraTS 2019 dataset and the results reported in Table \ref{BraTS2019}. Our TMA-TransBTS achieves DSC of 72.52\%, 78.73\%, 91.42\%, and HD of 9.138 mm, 9.401 mm, and 5.697 mm on ET, TC and WT. 
The pure CNN-based Residual 3D U-Net method achieves an average DSC of 78.47\%. 
Among existing hybrid CNN-Transformer based methods, Swin UNETR achieves the best average DSC of 80.01\%, and UNETR achieves the best average HD of 11.396 mm. 
Our TMA-TransBTS outperforms Residual 3D U-Net and Swin UNETR by achieving an average DSC of 80.89\%. 
Further, TMA-TransBTS obtains a reduction of 3.317 mm over UNETR in terms of average HD metric. 
Compared with previous CNN-based 3D methods and CNN-Transformer Hybrid 3D methods, our TMA-TransBTS achieves favorable performance in all metrics. 

\textbf{BraTS 2020.} Table \ref{BraTS2020} presents the comparison of the BraTS 2020 dataset. We report results on three brain tumor classes. Among existing CNN-based 3D methods, 3D U-Net and Residual 3D U-Net achieve a mean DSC of 80.57\% and 80.75\%, respectively. Among existing hybrid CNN-Transformer based methods, SwinUNETR achieve the best average DSC of 81.21\%. UNETR++ obtains a performance of 6.749 mm mean HD. In comparison, our TMA-TransBTS achieves improved performance with a mean DSC of 82.27\% and a mean HD of 5.681 mm. 

\begin{figure}[htb]
  \begin{center}
      \includegraphics[width=0.49\textwidth]{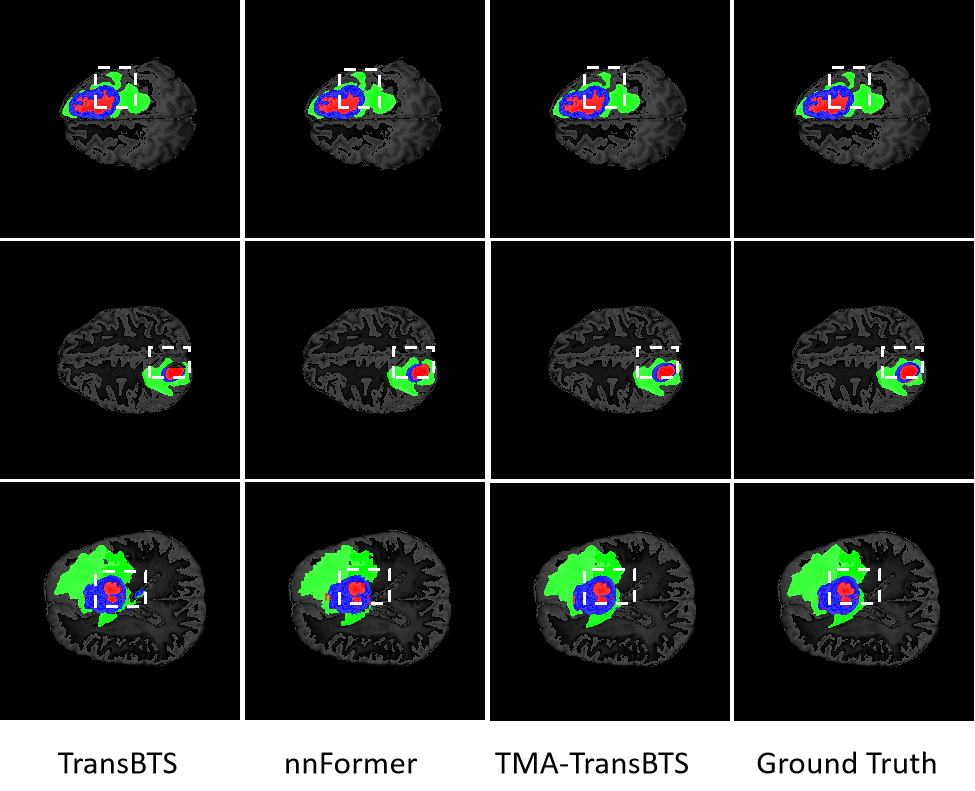}
      \caption{The visual comparison of MRI brain tumour segmentation results. The red regions denote the non-enhancing tumors, the green regions denote the peritumoral edema and the blue regions denote is enhancing tumors. }
      \label{Visual}
  \end{center}
\end{figure}

\begin{table}[htbp]
  \centering
  \caption{Investigation of the impact of different modules used in TMA-TransBTS on BraTS 2020. 'w/o' indicates no introduction of any of the proposed parts. '$+$ TMSM\_E' indicates the introduction of TMSM in the encoding phase. '$+$ TMSM\_D' indicates the introduction of TMSM in the decoding phase. '$+$ TMCM' indicates the introduction of TMCM. '$+$ DS' indicates the introduction of the deep supervision phase. It is worth noting that each line of the experimental method retains the part introduced in the previous line of the experimental method, and the last line of the experimental method is the complete model proposed in this chapter. Best results are bolded. }
    \begin{tabular}{l|c|c|c|c|c|c}
    \toprule
    \multicolumn{1}{c|}{\multirow{2}[3]{*}{Methods}} & \multicolumn{3}{c|}{DSC(\%)} & \multicolumn{3}{c}{HD(mm)} \\
\cmidrule{2-7}          & ET    & TC    & WT        & ET    & TC    & WT     \\
    \midrule
    w/o   & 54.92  & 65.43  & 76.85    & 23.030  & 31.970  & 40.850    \\
    + TMSM\_E & 67.85  & 76.28  & 89.26    & 6.890  & 8.530  & 5.480    \\
    + TMSM\_D & 70.69  & 78.46  & 89.24    & 6.430  & 7.100  & 5.770    \\
    + TMCM  & 73.38  & 79.03  & 92.08    & 6.320  & 8.500  & 4.910    \\
    + DS & \textbf{74.86}  & \textbf{79.64}  & \textbf{92.31}    & \textbf{5.900}  & \textbf{6.600}  & \textbf{4.540}    \\
    \bottomrule
    \end{tabular}%
  \label{Ablation}%
\end{table}%

\textbf{Visual results.} We show a visual comparison of the brain tumor segmentation results of various compared models including TransBTS\cite{TransBTS}, nnFormer\cite{nnFormer} and our TMA-TransBTS in Fig. \ref{Visual}. 
It is evident from Fig. \ref{Visual} that TMA-TransBTS can generate much better segmentation masks and describe brain tumors more accurately by extracting multi-scale features and modeling long-range dependencies among different modality volumes. 
We find that TMA-TransBTS maintains clear advantages over compared methods, one of which is that TMA-TransBTS is better at dealing with the brain tumor boundary.
For instance, in the fifth example of Fig. \ref{Visual}, TransBTS and nnFormer miss some important parts of the peritumoral edema compared to TMA-TransBTS. 

\subsection{Model Complexity}
Our TMA-TransBTS achieves the best segmentation results in all average metrics with 30.85M parameters and 141.79G FLOPs, which is a moderate-sized model. 
Compared with 3D U-Net, our TMA-TransBTS has more model parameters, but TMA-TransBTS achieves a 92.54\% reduction in FLOPs and significant improvements in all segmentation metrics. 
Compared with UNETR++, our TMA-TransBTS achieves 27.67\% reduction in model parameters (30.85M of TMA-TransBTS vs. 42.65M of UNETR++) and similar FLOPs (141.79G of TMA-TransBTS vs. 139.41G of UNETR++). 
In addition to 3D U-Net and UNETR++, our TMA-TransBTS has the minimum model size compared with other comparison methods. 

\subsection{Ablation Study}

Table \ref{Ablation} displays our ablation study results towards different modules in TMA-TransBTS. 
Introducing the TMSM within our TMA-TransBTS encoders leads to a significant improvement in performance over the method without any module introduction. 
The performance is further improved by integrating the TMSM in the decoding phase. 
This reveals the benefits of introducing TMSM in the encoding and decoding phases to extract 3D multi-scale features and establish long-range dependencies. 
Although the method that introduces TMSM in the decoding phase decreases the DSC metric for WT by 0.02\% and increases the HD metric for WT by 0.29 mm compared to introducing TMSM in the encoding phase only, it improves the average DSC metric by 5.00\% and decreases the average HD metrics by 1.60 mm. 
The phenomenon indicates that providing a multi-scale receptive field can be beneficial to the segmentation task. 
Afterwards, we use TMCM to replace traditional skip connection operations. 
Introducing TMCM brings a 6.10\% improvement in average DSC and a 0.43 mm reduction in average HD, which demonstrates that the TMCM may serve as a choice other than traditional skip connection operations. 
In addition to this, the introduction of a deep supervised phase can help to improve the stability of the model during the training phase and enhance the ability of the model to segment different levels of lesion regions. 

\section{Conclusion}
In this work, we propose a CNN-Transformer hybrid 3D medical image segmentation model, named TMA-TransBTS. 
Our TMA-TransBTS introduces a novel 3D Multi-scale Self-attention Module, which efficiently establishes long-range dependencies in 3D multi-modal data and captures rich multi-scale features. 
We also propose a 3D multi-scale cross-attention module to establish a link between the encoding stage and the decoding stage in TMA-TransBTS. 
Besides, we introduce a deep supervision stage during the model training stage to improve the ability of our TMA-TransBTS to learn deep features from 3D multi-modality medical images. 
Experimental results on three datasets (BraTS 2018, 2019, and 2020) validate the effectiveness of our TMA-TransBTS. 
In future work, we will develop a lightweight version of TMA-TransBTS and explore more efficient attention mechanisms in various Transformer variants for 3D multi-modality medical segmentation tasks. 

\section*{Acknowledgment}

This study was supported by Natural Science Foundation of Sichuan, China(No. 2023NSFSC0468, No. 2023NSFSC0031)


\bibliographystyle{IEEEtran}
\bibliography{IEEE-conference-template-062824.bib}

\end{document}